\documentclass[fleqn,twoside]{article}
\usepackage{amssymb}
\usepackage{amsbsy}
\usepackage{amsfonts}
\usepackage{epsfig}
\usepackage{espcrc2}

\usepackage{graphicx}
\usepackage{epsf}
\usepackage[figuresright]{rotating}

\newcommand{\eq}{\begin{equation}}
\newcommand{\ee}{\end{equation}}
\newcommand{\ea}{\begin{eqnarray}}
\newcommand{\eea}{\end{eqnarray}}
\newcommand{\be}{\begin{equation}}
\newcommand{\bea}{\begin{eqnarray}}

\newcommand{\AmS}{{\protect\the\textfont2
  A\kern-.1667em\lower.5ex\hbox{M}\kern-.125emS}}

\title{\vspace{-2.0cm}
       {\normalsize DESY 01--171}    \\[-0.2cm]
       {\normalsize KANAZAWA 01-15}   \\[0.850cm]
On the dynamics of color magnetic monopoles in full 
QCD\thanks{Talks given by H. Ichie, Y. Koma and T. Streuer at Lattice 2001,
Berlin, Germany.}}

\author{V. Bornyakov\address{NIC/DESY Zeuthen, Platanenallee 6, D-15738 Zeuthen,
Germany\\[-0.5em]}%
\thanks{On leave of absence from IHEP, Protvino, Russia.},
H. Ichie\address{Humboldt-Universit\"at zu Berlin, Institut 
f\"ur Physik, D-10115 Berlin, Germany\\[-0.5em]}, 
S. Kitahara\address{Institute for Theoretical Physics, Kanazawa University, Kanazawa 920-1192, Japan\\[-0.5em]}, 
Y. Koma$^{\rm c}$ ,  
Y. Mori$^{\rm c}$ ,                                    
Y. Nakamura$^{\rm c}$ ,                                    
M. Polikarpov\address{ITEP, B.Cheremushkinskaya 25, 
RU-117259 Moscow, Russia\\[-0.5em]},\\
G. Schierholz$^{\rm a,}$%
\address{Deutsches Elektronen-Synchrotron DESY,
D-22603 Hamburg, Germany\\[-0.5em]},
T. Streuer\address{Sektion Physik, Universit\"at 
M\"unchen, D-80333 M\"unchen,  Germany\\[-0.5em]},
H. St\"uben\address{Konrad-Zuse-Zentrum f\"ur Informationstechnik 
Berlin, D-14195 Berlin, Germany}, and 
T. Suzuki$^{\rm c}$ \\
}
 
\begin{document}

\begin{abstract}
We present first results on the dynamics of monopoles in full QCD with
$N_f=2$ flavors of dynamical quarks. Among the 
quantities being studied are the monopole density and the monopole 
screening length, the static potential as well as the profile of the color
electric flux tube. Furthermore, we derive the low-energy effective monopole 
action.
\vspace{1pc}
\end{abstract}

\maketitle

\section{INTRODUCTION}

In the dual superconductor picture of confinement the crucial degrees of 
freedom are the color magnetic monopoles revealed after abelian projection. 
In the maximally abelian gauge~\cite{klsw} one finds that the string tension 
is almost 
entirely due to monopole currents~\cite{monst,bbms}, and that the low-energy effective 
monopole 
action reproduces the string tension and the low-lying glueball masses 
quite well~\cite{ts}. Furthermore, the monopole currents show non-trivial 
correlations 
with gauge invariant excitations of the vacuum, such as the action and 
topological charge density~\cite{bcp,hart,bs}. The analysis of the monopole degrees 
of freedom
may thus provide 
important information about the confinement mechanism. Past studies have 
mainly been restricted to the quenched approximation. It will be interesting 
now to see how the vacuum reacts to the presence of dynamical color electric
charges.

Our studies will be done on $N_f=2$ dyna\-mi\-cal gauge field configurations 
generated by the QCDSF and UKQCD collaborations using non-perturbatively 
$O(a)$ improved Wilson fermions~\cite{lambda}:
\begin{displaymath}
S_F = S^{(0)}_F - \frac{\rm i}{2} \kappa\, g\, 
c_{SW} a^5 
\sum_x \bar{\psi}(x)\sigma_{\mu\nu}F_{\mu\nu}\psi(x),
\end{displaymath}
where $S^{(0)}_F$ is the original Wilson action. 

The link variables $U(s,\mu)$ are brought into the maximally abelian 
gauge by maximizing the quantity~\cite{bsw}
\begin{equation}
R=\sum_{s,\mu}\,\sum_{i=1}^3 |\widetilde{U}_{ii}(s,\mu)|^2
\end{equation}
with respect to gauge transformations $g$:
\begin{equation}
\widetilde{U}(s,\mu)=g(s)U(s,\mu)g(s+\hat{\mu})^{-1},
\end{equation}
using an $SU(3)$ update of the simulated annealing algorithm first introduced
in~\cite{bbmp} for $SU(2)$. The abelian link variables are defined by
\begin{equation}
u_i(s,\mu) \equiv e^{i\theta_i(s,\mu)}
\label{abel}
\end{equation}
with
\begin{eqnarray}
\theta_i(s,\mu) &=& \arg \widetilde{U}_{ii}(s,\mu) \nonumber \\
&-&\frac{1}{3}\sum_{j=1}^3 \arg \widetilde{U}_{jj}(s,\mu)|_{{\rm mod}\,
2\pi} .
\end{eqnarray}

\section{MONOPOLE DENSITY}

The first quantity we looked at is the monopole density. This is defined by
\begin{equation}
\rho=\frac{1}{12V}\sum_{s,\mu} \sum_{i=1}^3 |k_i (s,\mu)|
\end{equation}
where $k_i(s,\mu)$ are the monopole currents, which are obtained from the 
angles $\theta_i(s,\mu)$ \cite{dt}, and $V$ is the lattice
volume. The currents are subjected to the constraint $\sum_i k_i(s,\mu)=0$.
\begin{figure}[t!b]
\hbox{
\epsfysize=5.0cm
\epsfxsize=7.cm
\epsfbox{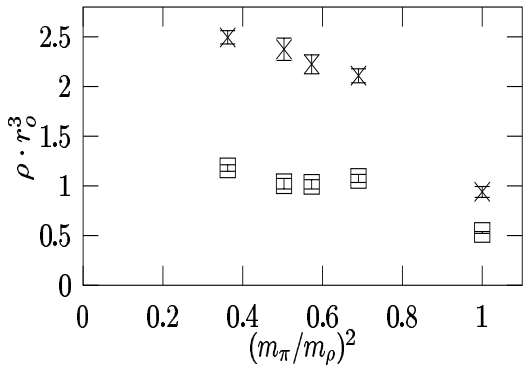}}
\vspace{-.5cm}
\caption{\it The monopole density for all clusters $(\times)$ and for large 
clusters $(\square)$.}
\label{rho_fig}
\end{figure}
The monopole currents fall into clusters of different lengths:
large clusters are responsible for quark confinement, while short 
clusters are mainly due to ultraviolet fluctuations and have no effect on the 
confining forces \cite{ht}. As large clusters we count the largest cluster of the
system and all clusters that wind around the lattice at least once. It was
shown~\cite{bm} (in pure $SU(2)$ gauge theory) that the density of 
monopoles belonging to the large clusters is finite in the continuum limit.

In Fig.~1 we show the monopole density for all clusters, and separately for 
the large clusters, as a function of $(m_\pi/m_\rho)^2$. Also shown is the 
quenched result ($m_\pi/m_\rho = 1$). The quenched data set, $\beta = 6.0$,
was chosen to match the lattice spacing of the dynamical configurations, 
$a \approx 0.1$ fm. The lattice volume varies between $16^3 32$ and $24^3 48$.
We see that the monopole density
is strongly affected by the presence of dynamical quarks. On the dynamical
configurations the density is twice as large as on the quenched configurations
with increasing tendency towards smaller quark masses.

\section{STATIC POTENTIAL}

\begin{figure}[t!]
\vspace{.1cm}
\hbox{
\epsfysize=5.0cm
\epsfxsize=7.cm
\epsfbox{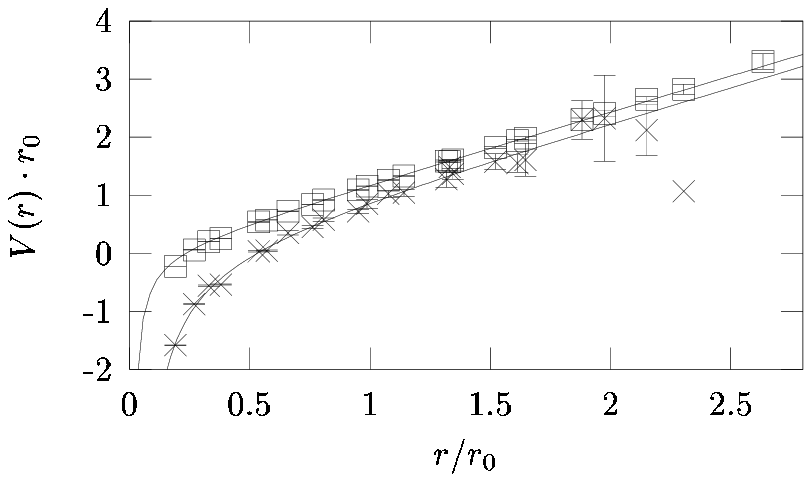}}
\hbox{
\epsfysize=5.0cm
\epsfxsize=7.cm
\epsfbox{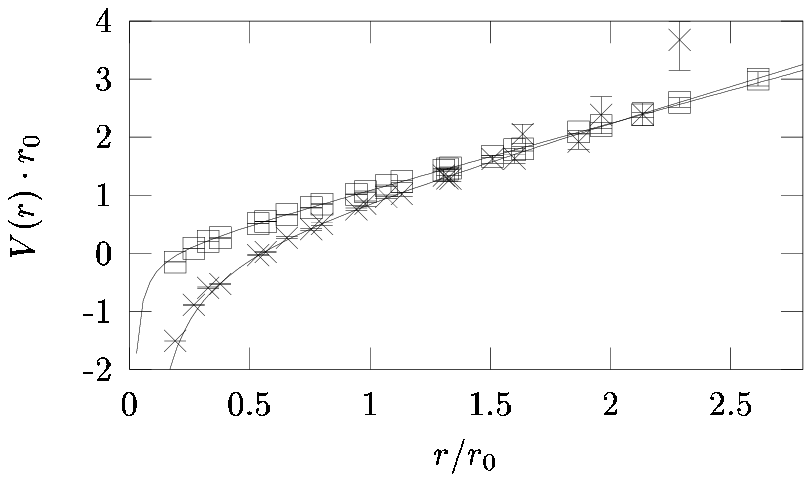}}
\vspace{-.9cm}
\caption{\it The static potential $(\times)$ and the abelian static potential 
$(\square)$, with the self-energy subtracted. The upper figure is for
$\beta=5.29, \kappa=0.135$, the lower figure for $\beta=6.0$ in the quenched
theory.}
\vspace{-0.65cm}
\label{pot1}
\end{figure}
The next quantity we have looked at is the static potential $V(r)$. This is
computed from smeared Wilson loops. We parameterize the potential by
\begin{equation}
V(r) = V_0 -\frac{e}{r} + \sigma\, r.
\end{equation}
We also have looked at the abelian potential built from abelian link 
variables (\ref{abel}). This defines an abelian string tension 
$\sigma^{\rm ab}$. In Fig.~2 we show $V(r)$ and its abelian counterpart, and 
compare it with the quenched result. A fit gives 
$\sigma^{\rm ab}/\sigma = 0.94(8)$ 
for the dynamical configuration shown and 
$\sigma^{\rm ab}/\sigma = 1.03(15)$ for the quenched case.

\begin{figure}[t!b]
\hbox{
\epsfysize=5.0cm
\epsfxsize=7.cm
\epsfbox{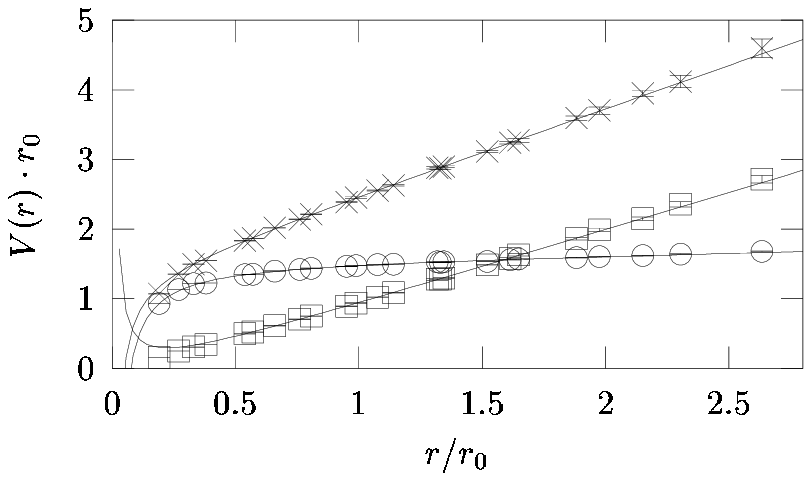}}
\hbox{
\epsfysize=5.0cm
\epsfxsize=7.cm
\epsfbox{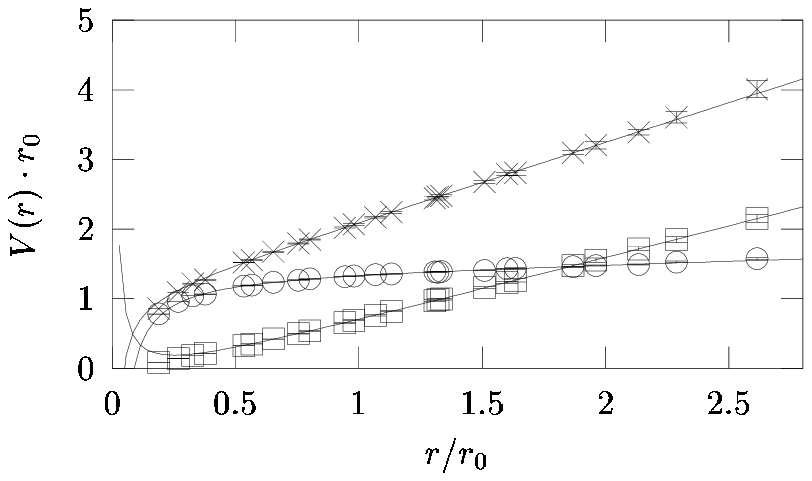}}
\vspace{-.5cm}
\caption{\it The abelian static potential $(\times)$ divided into monopole 
$(\square)$ and photon $(${\LARGE $\circ$}$)$ contributions. The upper figure is 
for $\beta=5.29, \kappa=0.135$, the lower figure for $\beta=6.0$ in the 
quenched theory.}
\vspace{-0.5cm}
\label{pot2}
\end{figure}
It is instructive to separate the abelian gauge potential into a monopole
and `photon' part, $\theta(s,\mu) = \theta^{\rm mon}(s,\mu) +
\theta^{\rm ph}(s,\mu)$ where
\begin{equation}
\theta^{\rm mon}(s,\mu) = -2\pi\sum_{s^\prime}D(s-s^\prime) \nabla_\mu^- 
m_{\mu\nu}(s^\prime).
\end{equation}
Here $D(s)$ is the Coulomb propagator and $m_{\mu\nu}(s)$ is the number of 
Dirac strings piercing the abelian plaquette $u(s,\mu,\nu)$.
In Fig.~3 we show the abelian potential constructed from the monopole 
and `photon' part of the abelian gauge potential, respectively. We see that
most of the string tension is due to the monopole currents, while the `photon' 
part of the potential is Coulomb-like, exactly as in the quenched theory.

\begin{figure}[t!]
\hbox{
\epsfysize=5.0cm
\epsfxsize=7.cm
\epsfbox{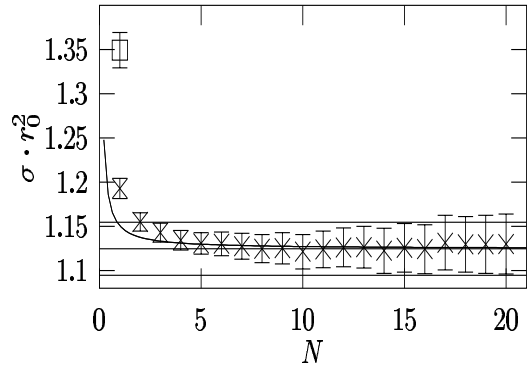}}
\vspace{-.5cm}
\caption{\it The abelian string tension versus the number of gauge copies 
$(\times)$, and for iterative gauge fixing $(\square)$.}
\label{st_gc}
\vspace{-0.5cm}
\end{figure}

The quantity $R$ may have many local maxima. Numerically it is difficult
to get to the absolute maximum, which constitutes a problem, the Gribov
problem. In this study we considered a single gauge copy per configuration. 
To estimate the error arising from incomplete gauge fixing we 
created a sequence of random gauge copies on each of our configurations at 
$\beta=5.29, \kappa=0.135$, following~\cite{bbms}. In Fig.~4 we show the 
abelian string tension
as a function of the number of gauge copies considered (thereby always
computing $\sigma^{\rm ab}$ on the copy with the largest value of $R$). 
We also show the result for iterative gauge fixing. We see that the bias
induced by incomplete gauge fixing is about 6\%, while iterative
gauge fixing may give results that are wrong by $O(20\%)$. Therefore careful
gauge fixing is necessary. Similar results hold for other observables.

\section{SCREENING LENGTH}

The screening length $\xi$ of a plasma of monopoles is defined by the 
exponential 
decrease of the magnetic flux through a sphere of radius $r$ around the 
monopole:
\begin{equation}
\Phi(r)=\Phi_0 e^{-r/\xi}.
\end{equation} 
A screened plasma of static monopoles is known to generate a string 
tension~\cite{hart} 
\begin{equation}
\sigma \propto \rho\, \xi.
\label{xi}
\end{equation}
If this is true for dynamical monopoles as well, we would expect
the screening length to be significantly smaller on dynamical 
configurations.
Considering the monopoles in the large clusters only, we find
\begin{displaymath}
\begin{tabular}{c|c|c}
$m_\pi/m_\rho$ & $\xi/r_0$ & $\sigma^{\rm ab}/\rho\, \xi$ \\
\hline
0.602 & 0.484(19) & 0.49(4) \\
0.709 & 0.466(26) & 0.38(4) \\
0.757 & 0.521(17) & 0.43(3) \\
0.831 & 0.482(17) & 0.40(3) \\
1     & 0.662(34) & 0.31(3) 
\end{tabular}
\end{displaymath}
We see that the screening length is indeed $O(40\%)$ lower in the dynamical 
case. We interprete this, as well as the increase in the monopole density, as 
pairing effect~\cite{ah}.

\section{ABELIAN FLUX TUBE}

The chromoelectric flux tube between static color electric charges forms 
an interface between the superconducting (confining) phase and the conducting 
phase, and in this capacity bears valuable information about the
properties of the QCD vacuum. Because we are primarily interested in the
dynamics of monopoles, we may restrict ourselves to abelian 
flux tubes being excited by abelian Wilson loops. The abelian Wilson loop is 
defined by
\begin{equation}
W(R,T) = \frac{1}{3}\,\mbox{Tr}\, \prod_{s \in {\mathcal C}} u(s,\mu)
=\frac{1}{3}\,\mbox{Tr}\,e^{{\rm i} \theta_{\mathcal C}},
\end{equation}
where ${\mathcal C}$ is a rectangular loop of extension $R \times T$,
$\theta_{\mathcal C} = 
\mbox{diag}[\theta_1,\theta_2,\theta_3]_{\mathcal C}$.
In this section we shall study the spatial profile of the flux tube. This is 
done by looking at correlations of appropriate operators ${\mathcal O} = 
\mbox{diag}[{\mathcal O}_1,{\mathcal O}_2,{\mathcal O}_3]$ with
the abelian Wilson loop operator:
\begin{equation}
\frac{1}{3} \frac{\langle\mbox{Tr}\,{\mathcal O}(s) \,\mbox{Tr}\,
e^{{\rm i} \theta_{\mathcal C}}\rangle}{\langle\mbox{Tr}\,
e^{{\rm i} \theta_{\mathcal C}}\rangle} -
\frac{1}{3} \langle \mbox{Tr}{\mathcal O}\rangle 
\label{ceven}
\end{equation}
for C-parity even operators, and 
\begin{equation}
\frac{\langle\mbox{Tr}\,({\mathcal O}(s)\, 
e^{{\rm i} \theta_{\mathcal C}})\rangle}{\langle\mbox{Tr}\,
e^{{\rm i} \theta_{\mathcal C}}\rangle}  
\end{equation}
\begin{figure}[t!hb]
\vspace{0.0cm}
\hbox{
\hspace{0.1cm}
\epsfxsize = 6.5 cm
\epsfysize = 4.3 cm
\epsfbox{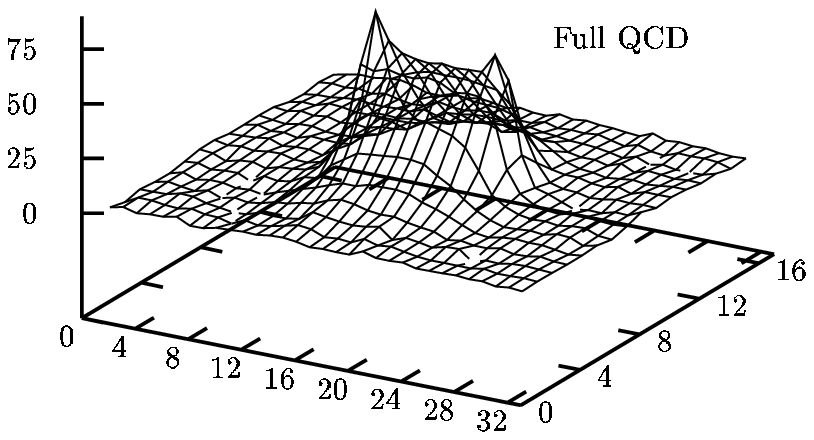}
}
\vspace{0.2cm}
\hbox{
\hspace{0.1cm}
\epsfxsize = 6.5 cm
\epsfysize = 4.3 cm
\epsfbox{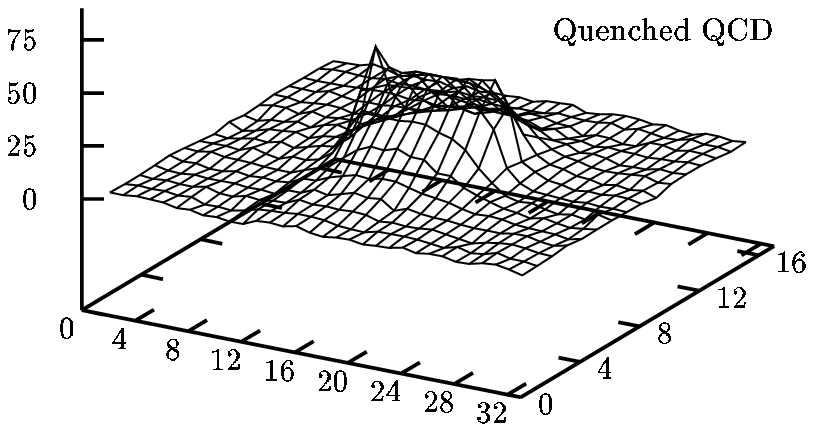}
}
\vspace{-0.7cm}
\caption{\it The action density $\rho_A(s) r_0^4$ of the abelian flux tube 
in full QCD ($m_\pi/m_\rho \approx 0.6$) and quenched QCD.}
\label{action}
\end{figure}
\begin{figure}[t]
\vspace{0.cm}
\hbox{
\hspace{0.4cm}
\epsfxsize = 6.2 cm
\epsfysize = 4.1cm
\epsfbox{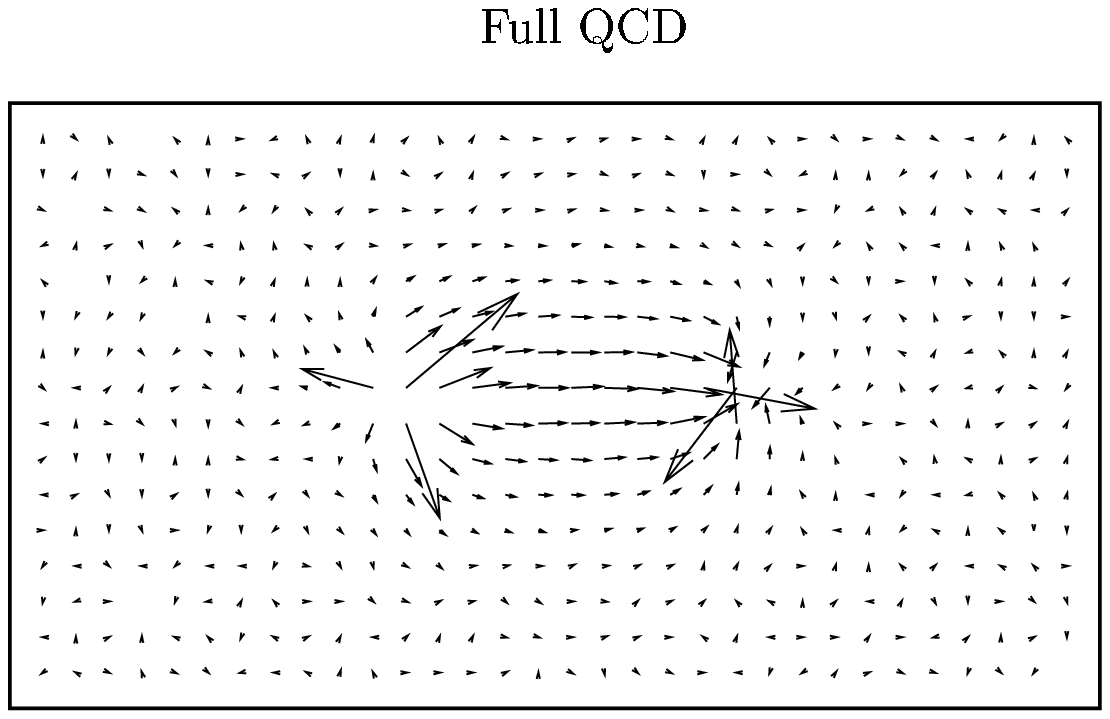}
}
\vspace{0.5cm}
\hbox{
\hspace{0.4cm}
\epsfxsize = 6.2 cm
\epsfysize = 4.1cm
\epsfbox{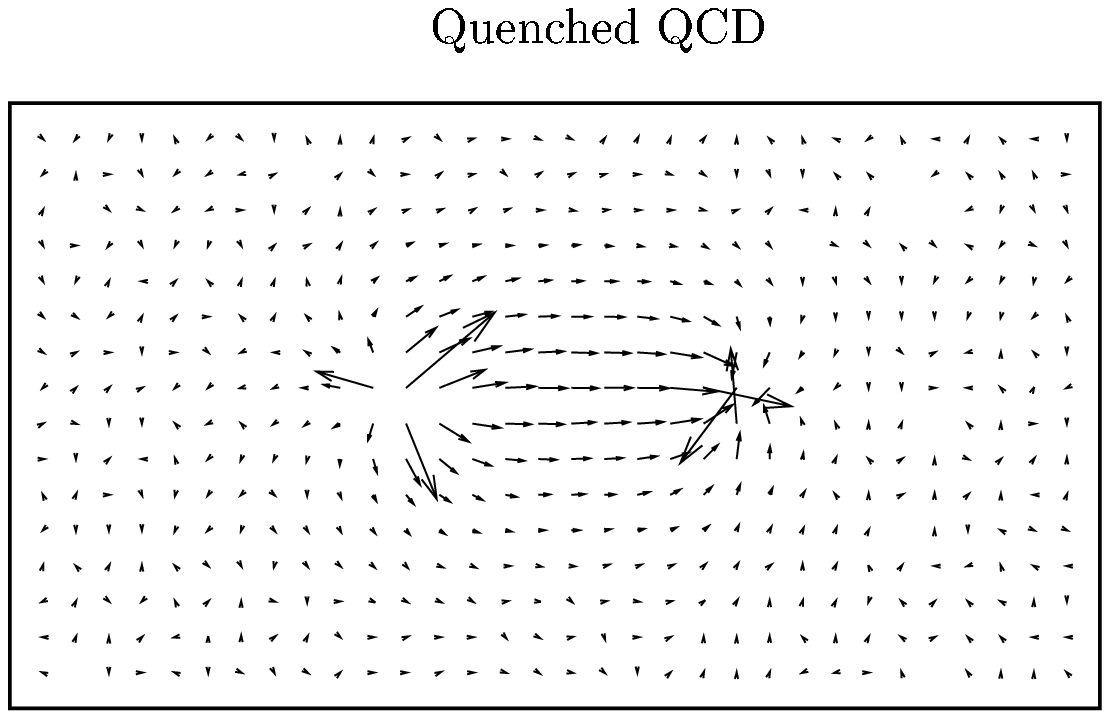}
}
\vspace{-.6cm}
\caption{\it Distribution of the color electric field.}
\vspace{-.4cm}
\label{fluxtube1}
\end{figure}
for C-parity odd operators, 
where $s$ is taken to lie on the $R \times T$
plane at $T/2$.

The first quantities we looked at are the action density 
\be
\mbox{Tr}{\mathcal O}(s) =  \frac{\beta}{3} \sum_{\mu>\nu,i} 
\cos \theta_i(s,\mu, \nu). 
\ee
and the color electric field
\begin{equation}
{\mathcal O}_i(s) = E_j^i(s) = \mbox{i}\bar{\theta}_i(s,4,j),
\end{equation}
where
\be
\bar{\theta}_i(s,\mu, \nu) = -\mbox{i} \ln(e^{i\theta_i(s,\mu, \nu)}).
\ee
In Figs.~5 and 6 we show our results on a $16^3 32$ lattice, where we took
$R=10$, which corresponds to a separation of $\approx 1$ fm, and $T=6$. 
The action density is somewhat larger
in the dynamical case, while we see little difference in the distribution
of the electric field between full and quenched QCD. A fit of $E^i_\| = 
\mbox{const.}\,\exp(-r_\perp/\lambda)$, restricting $r_\perp$ to $r_\perp > 
0.25$ fm, gives the penetration length $\lambda = 0.15(2)$ fm in full QCD 
and $0.17(1)$ fm in the quenched theory.

\begin{figure}[t!h]
\vspace{0.0cm}
\hbox{
\hspace{0.1cm}
\epsfxsize = 6.5 cm
\epsfysize = 4.3 cm
\epsfbox{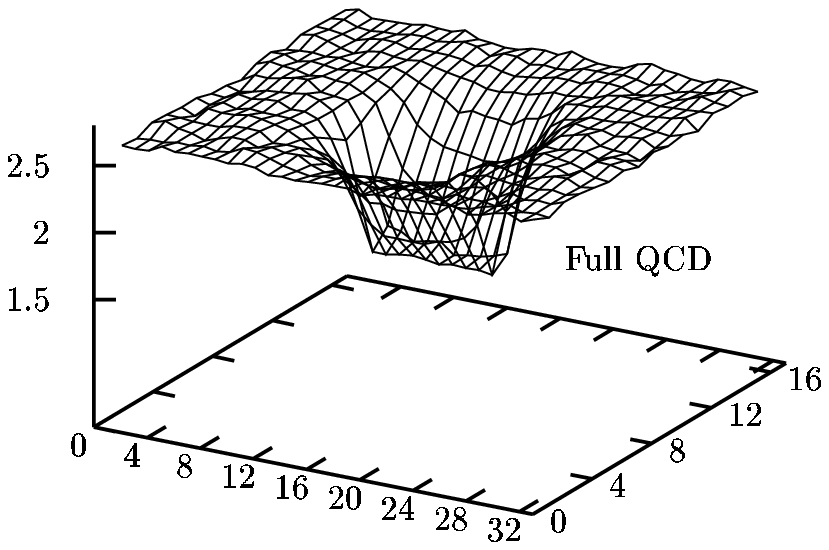}
}
\vspace{0.5cm}
\hbox{
\hspace{0.1cm}
\epsfxsize = 6.5 cm
\epsfysize = 4.3 cm
\epsfbox{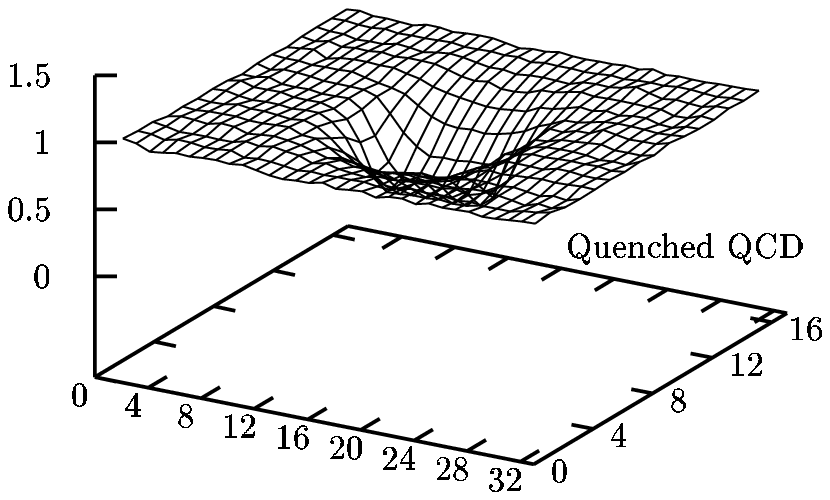}
}
\vspace{-.5cm}
\caption{\it The local monopole density $\rho(s)\, r_0^3$ in the vicinity 
of the abelian flux tube.}
\label{mondens}
\vspace{-.2cm}
\end{figure}
\begin{figure}[b!h]
\vspace{0.cm}
\epsfxsize = 7.6 cm
\epsfbox{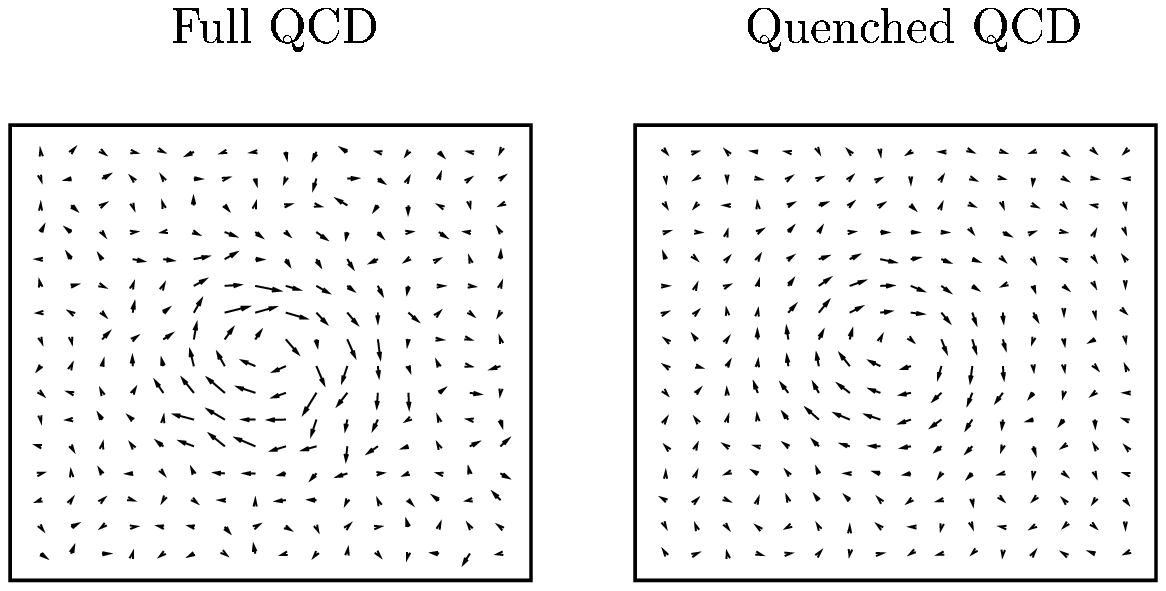}
\vspace{-0.5cm}
\caption{\it The solenoidal monopole current in a plane perpendicular to the 
abelian flux-tube, positioned at $R/2$.}
\label{monocur}
\end{figure}

Next we looked at the local monopole density 
\begin{equation}
\mbox{Tr}{\mathcal O}(s) = \frac{1}{12}\sum_{\mu,i}|k_i(s,\mu)|. 
\end{equation}
The result is shown in Fig.~7. In this case 
$(1/3) \langle\mbox{Tr}\,\mathcal O \rangle$
was not subtracted in eq.~(\ref{ceven}). This figure shows again that the 
monopole 
density, outside the flux tube, is more than twice as high on the dynamical 
configurations as in the quenched case. Inside the flux tube the 
monopole currents are strongly suppressed, as expected. Due to the high
monopole density outside the flux tube, the interface appears to be much 
`harder' in full QCD.

We expect the monopole current to form a solenoidal supercurrent which, 
similar to a coil in electrodynamics, constricts the color electric fields
into flux tubes. This is expressed by the dual Amp\`ere law
\begin{equation}
{\mathbf k} = {\mathbf \nabla} \times {\mathbf E}.
\label{ampere}
\end{equation}
We find that the monopole current is indeed azimuthal, as can be seen in
Fig.~8. In Fig.~9 we compare the l.h.s of eq.~(\ref{ampere}) with the r.h.s.
and find that Amp\`ere's law is very well satisfied on our dynamical 
configurations. Amp\`ere's law was already verified in pure $SU(2)$ gauge
theory in~\cite{Bali}.

\begin{figure}[!t]
\vspace{0.1cm}
\epsfxsize = 7 cm
\epsfbox{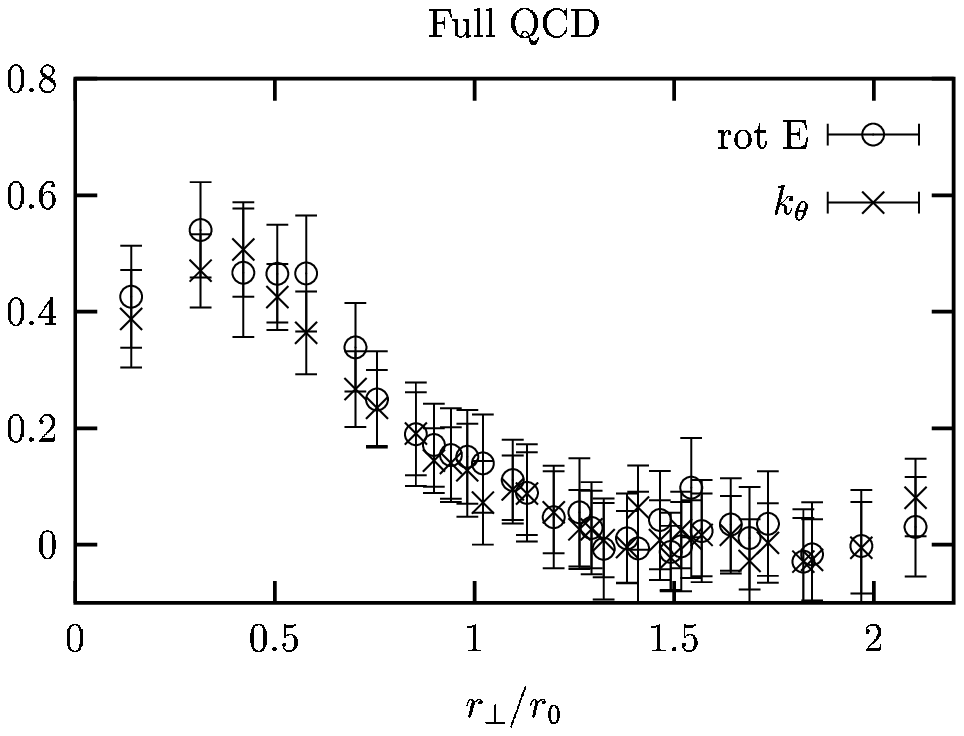}
\vspace{0.4cm}
\epsfxsize = 7 cm
\epsfbox{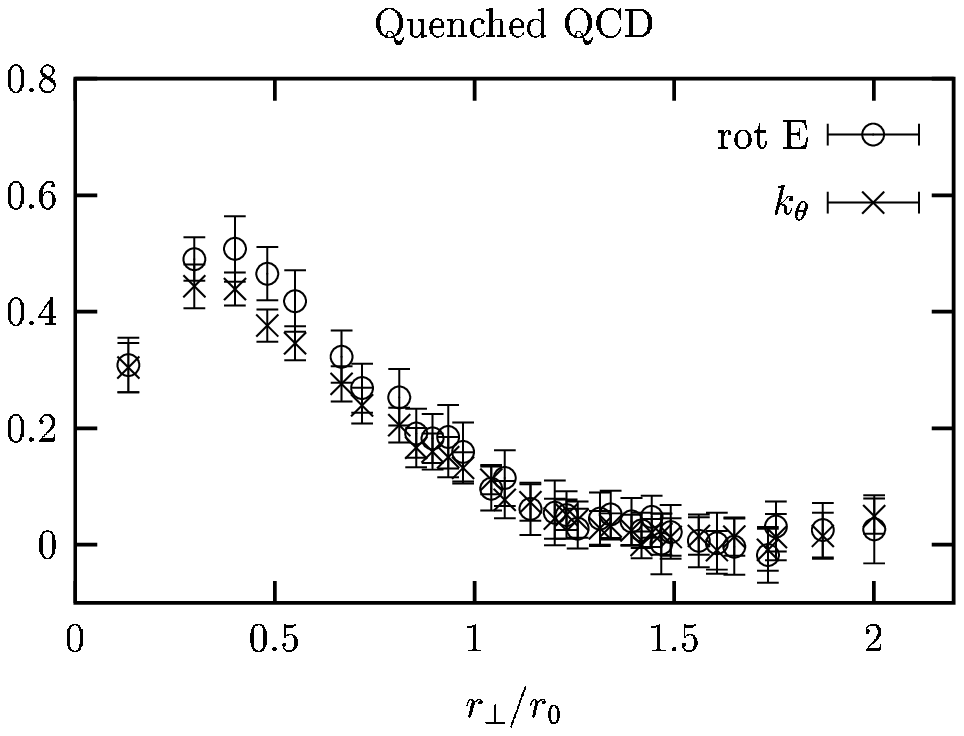}
\vspace{-0.75cm}
\caption{\it Comparing ${\mathbf k}$ and ${\mathbf \nabla} \times 
{\mathbf E}$.}
\label{amperef}
\vspace{-0.60cm}
\end{figure}
We expect the flux tube (string) to break eventually if the static charges are
pulled apart beyond a certain distance. That distance is expected to be 
between 1.5 and 2 fm, depending on the mass of the quark. In Fig.~10 we show 
the distribution of the color electric field in the vicinity of the flux tube
on the $24^3 48$ lattice at $\beta = 5.29, \kappa=0.1355$, corresponding to
an $m_\pi/m_\rho$ ratio of $ \approx 0.7$, for $R=22$ which amounts
to a separation of $\approx 2$ fm. We see no sign of string breaking yet. We
notice though that the electric field vectors show some level of noise half 
way between the charges.
\begin{figure}[t!b]
\hspace{.2cm}
\epsfxsize = 6.7 cm
\epsfbox{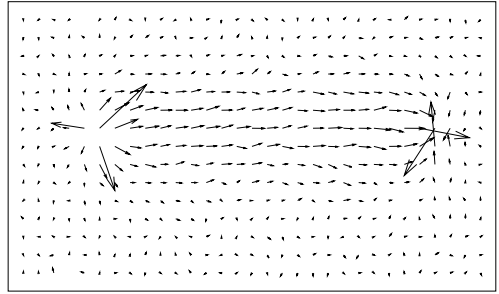}
\vspace{-0.5cm}
\caption{\it Distribution of the color electric field on the $24^3 48$ lattice 
for $R=22, T=5$ in full QCD.}
\label{fluxtube2}
\vspace{-0.75cm}
\end{figure}

The penetration length and half width of the flux tube turn out to be 
$\approx 0.16$ fm and $\approx 0.3$ fm, respectively. It is striking that the 
abelian flux tube does not show any broadening towards larger distances from 
the static charges, as one would expect if the fluctuations of the string were
described effectively by the Nambu-Goto action~\cite{lm}. (This effect has 
also been observed in pure $SU(2)$ gauge theory~\cite{bali}.) This could mean 
that the 
abelian string is described by the Ramond string action instead, which, for
example, does not show such a broadening effect~\cite{Olesen}.

\section{EFFECTIVE MONOPOLE ACTION}

We have seen that the vacuum undergoes several changes if dynamical color
electric charges are introduced. We shall study now how this will affect
the effective monopole action.

There are three types of monopole currents, of which two are independent. 
For simplicity we take into account only one of them, thus integrating
out the other two \cite{aekms}.
For the time being, we assume the form of the effective monopole action in 
full QCD to be the same as in the quenched theory~\cite{Yamagishi}. 
This is composed of 27 types of two-point interactions, one four-point 
interaction and one six-point interaction:
\begin{equation}
S(k)=\sum_{i=1}^{29} G_i S_i (k),
\end{equation}
where $G_i$ are the coupling constants which need to be determined. This
we will do by employing an extended Swendsen method~\cite{Shiba:1995pu}.
Explicitly, we have: 

\noindent
{\it Two-point interaction} ($i=1, \dots, 27$) 
\begin{eqnarray}
S_1 (k) &=& \sum_s\sum_{\mu =1}^4 k(s,\mu)^2,
\nonumber\\
S_2 (k) &=& \sum_s\sum_{\mu =1}^4 k(s,\mu)\,k(s+\hat{\mu},\mu),
\nonumber\\
S_3 (k) &=& \sum_s\sum_{\mu \ne \nu} k(s,\mu)k(s+\hat{\nu},\mu); \\
        &\vdots& \nonumber
\end{eqnarray}

\noindent
{\it Four-point interaction} 
\begin{equation}
S_{28} (k) = \sum_s \left( \sum_{\mu=-4}^4k(s,\mu)^2 \right)^2;
\end{equation}

\noindent
{\it Six-point interaction} 
\begin{equation}
S_{29} (k) = \sum_{s} \left( \sum_{\mu=-4}^4 k(s,\mu)^2 \right)^3.
\end{equation}

\begin{figure}[b!]
\begin{center}
\epsfysize=63mm
\epsfxsize=71mm
\vspace{-0.5cm}
\epsfbox{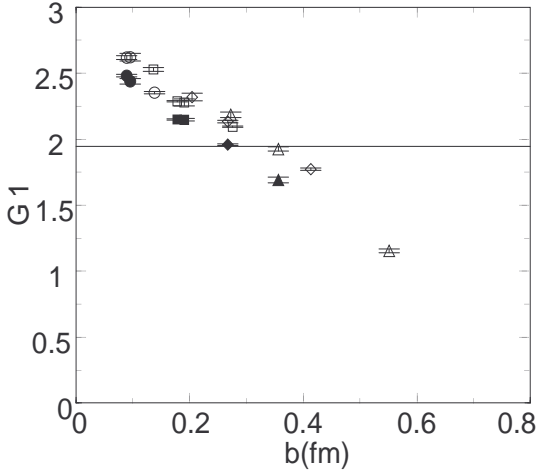}
\end{center} 
\vspace{-1.3cm}
\caption{\it The coupling constant of the monopole self-interaction, $G_1$, 
as a function of the physical length scale $b$. The symbols are: $n=1$
$(${\Large $\bullet$}$)$, $n=2$ $(\blacksquare)$, $n=3$ $(\blacklozenge)$,
$n=4$ $(\blacktriangle)$ for full QCD, and $n=1$
$(${\Large $\circ$}$)$, $n=2$ $(\square)$, $n=3$ $(\lozenge)$,
$n=4$ $(\triangle)$ for quenched QCD.}
\label{fig:g1-vs-b}
\vspace*{-0.5cm}
\end{figure}

Our calculations are done on $O(50)$ gauge fixed configurations each on the
$24^3 48$ lattice at $\beta=5.29, \kappa=0.1355$ and on the $16^3 32$ lattice 
at $\beta=5.29, \kappa=0.1350$~\cite{lambda}. For comparison, we have also
done calculations in the quenched theory on $24^3 48$ lattices at $\beta=5.8,
6.0$ and $6.2$, and on a $16^3 32$ lattice at $\beta = 6.0$. 

\begin{figure}[t!]
\begin{center}
\epsfysize=63mm
\epsfxsize=72mm
\vspace{0.1cm}
\epsfbox{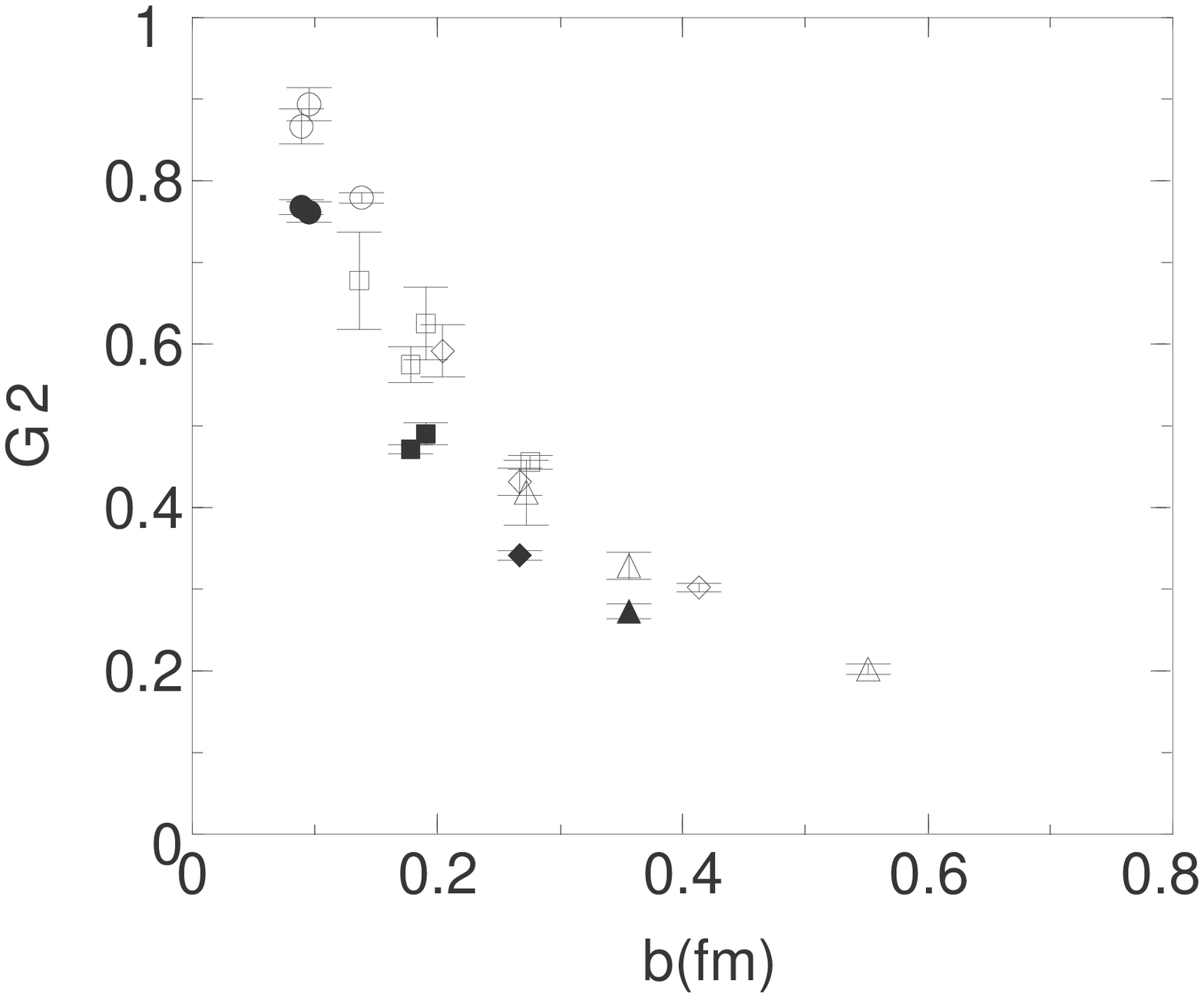}
\epsfysize=63mm
\epsfxsize=72mm
\epsfbox{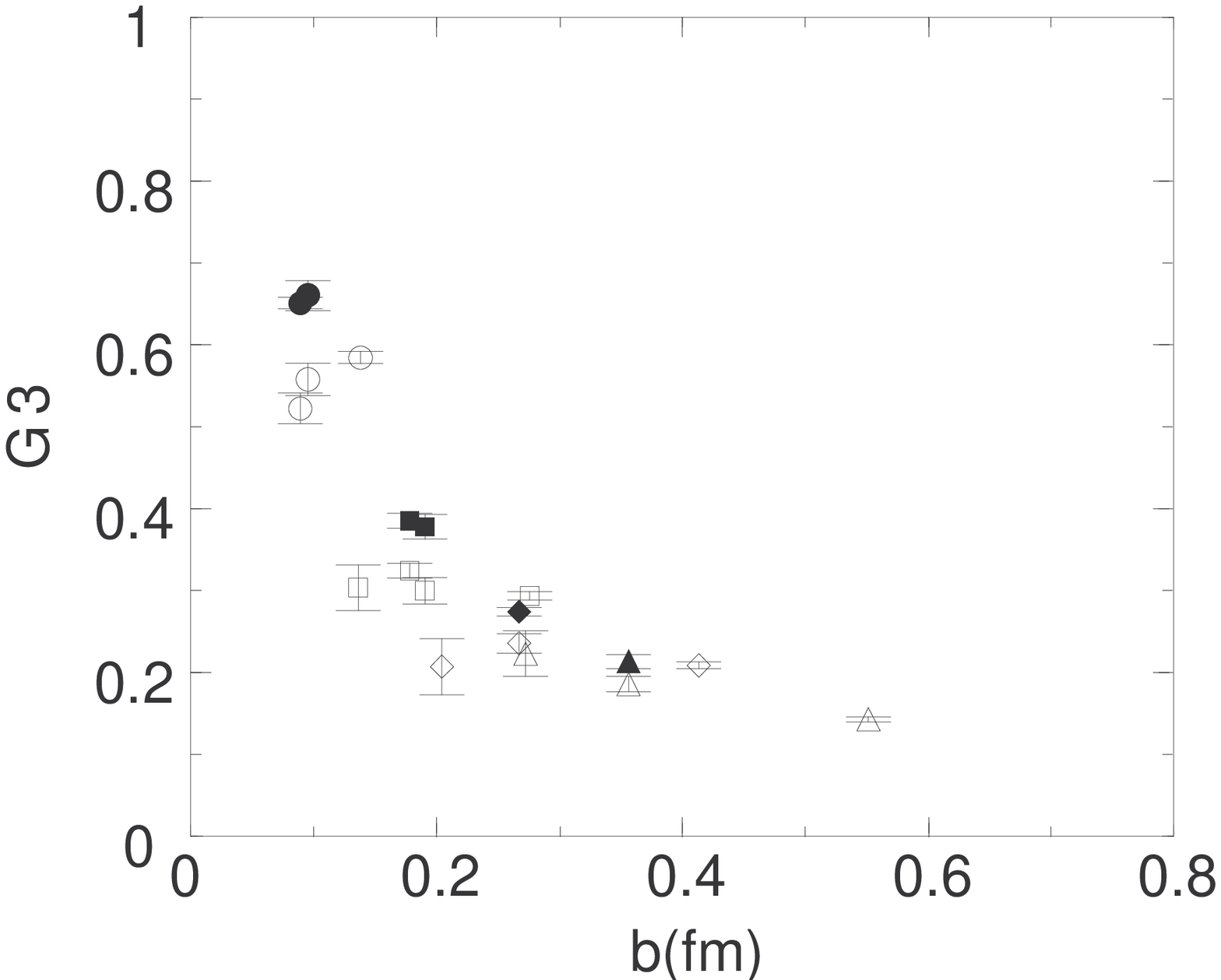}
\end{center} 
\vspace{-1.2cm}
\caption{\it The coupling constants $G_2$ and $G_3$ 
as a function of $b$. The symbols are as in Fig.~11.}
\label{fig:g2-g3-vs-b}
\vspace*{-0.5cm}
\end{figure}

\begin{figure}[h!]
\begin{center}
\epsfysize=63mm
\epsfxsize=72mm
\vspace{0.1cm}
\epsfbox{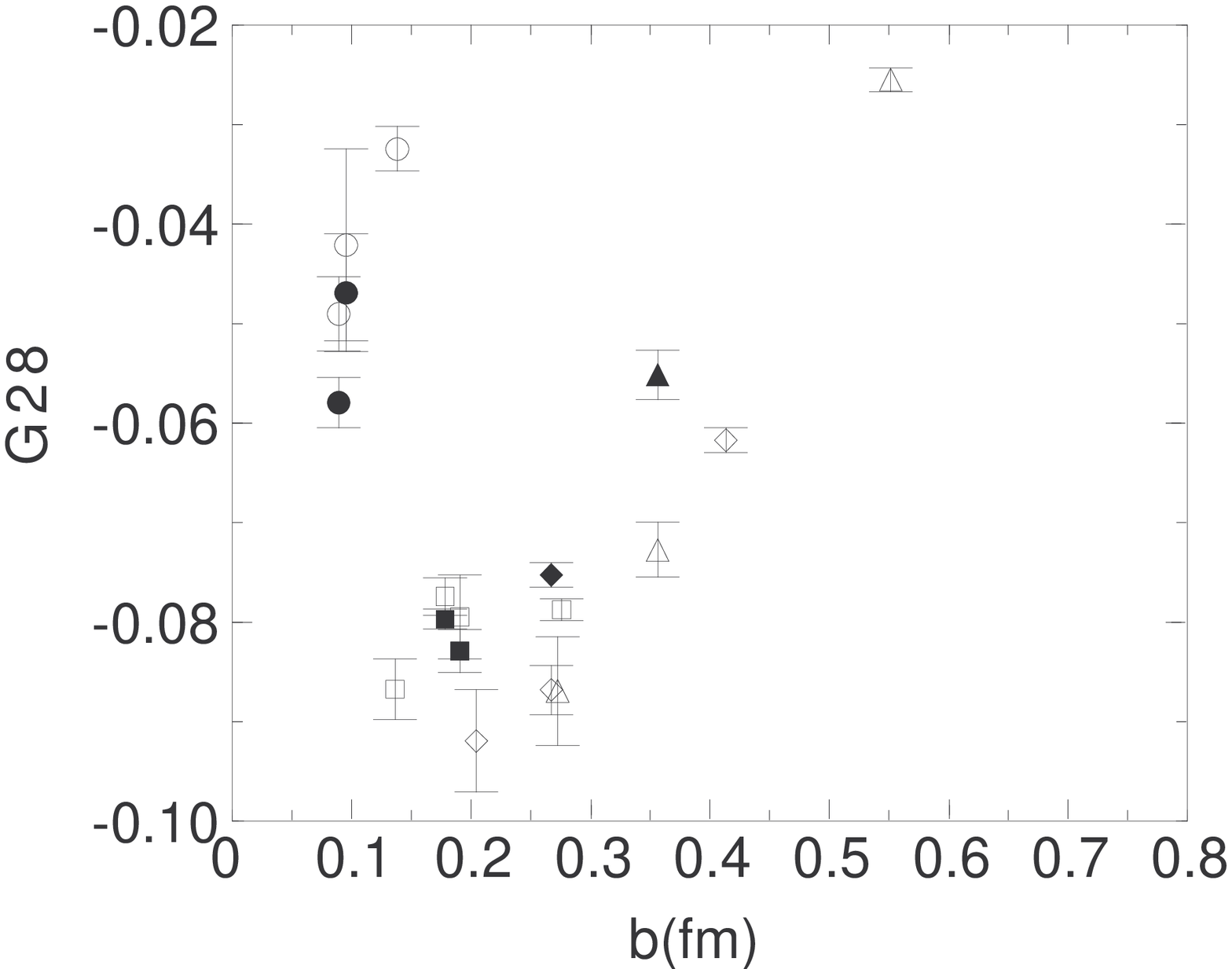}
\epsfysize=63mm
\epsfxsize=72mm
\epsfbox{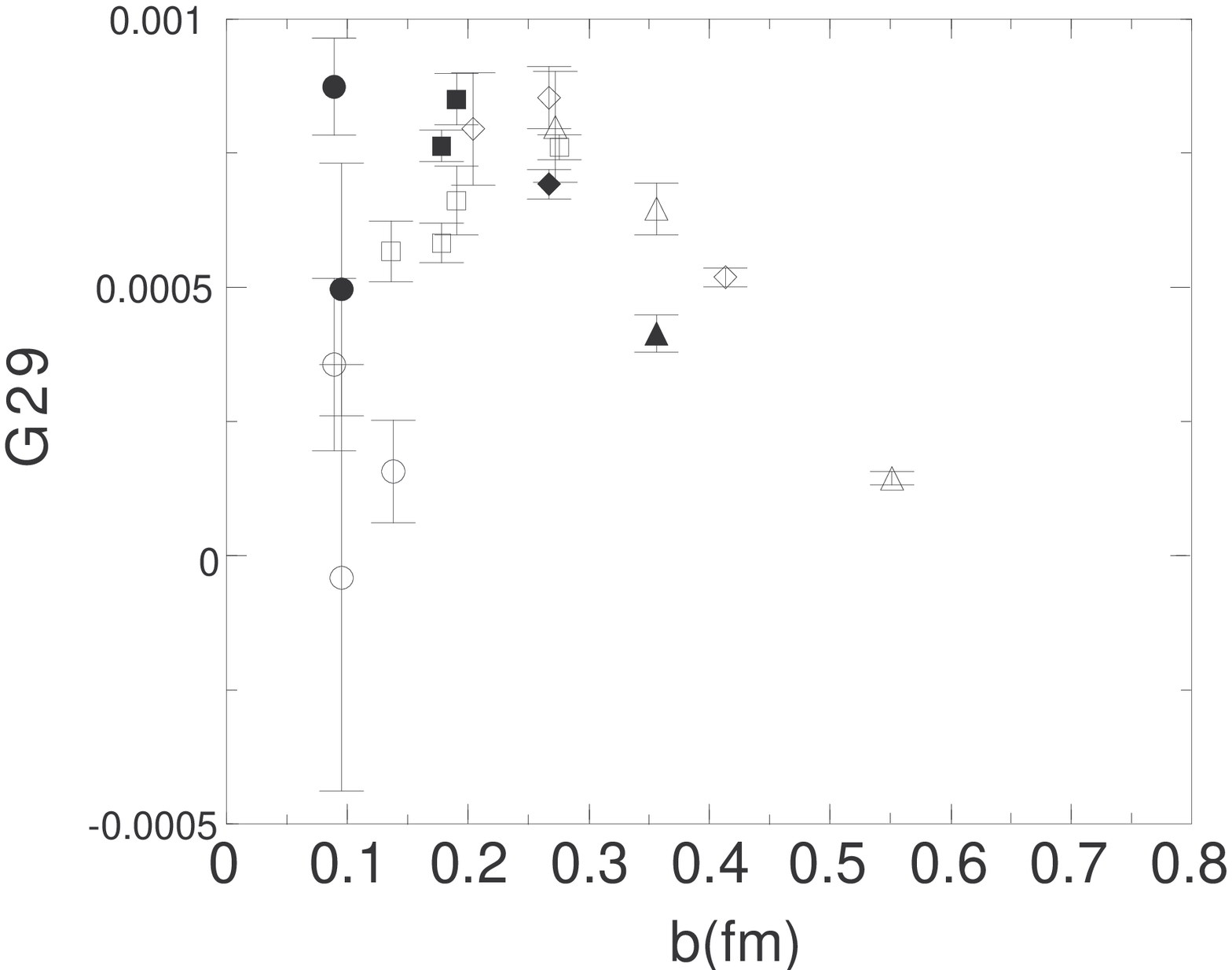}
\end{center} 
\vspace{-1.2cm}
\caption{\it The coupling constants $G_{28}$ and $G_{29}$ of the four-point and 
six-point interactions, respectively, as a function of $b$. The symbols are 
as in Fig.~11.}
\label{fig:g28-g29-vs-b}
\vspace*{-0.5cm}
\end{figure}

We have employed a type-II block spin transformation \cite{Ivanenko}
with up to $n-1=3$ blocking steps. 
In Fig.~\ref{fig:g1-vs-b} we show the self-coupling $G_1$ as a 
function of the physical length scale $b=na$ for both full and quenched QCD.
We see that $G_1^{\rm full}$ is systematically smaller than $G_1^{\rm 
quenched}$ for all $b$, in agreement with the higher monopole density
in full QCD. A necessary condition for monopole condensation is 
$G_1 \leq \ln 7$. This is achieved for $b \gtrapprox 0.27$ fm in full QCD and 
for $b \gtrapprox 0.35$ fm in the quenched theory. In 
Fig.~\ref{fig:g2-g3-vs-b} we show the coupling constants $G_2$ and $G_3$.
We see that $G_2^{\rm full} < G_2^{\rm quenched}$ as in the previous case, 
while $G_3$ shows the opposite behavior: 
$G_3^{\rm full} > G(3)^{\rm quenched}$. For the other coupling constants we 
find little
difference between full QCD and the quenched theory. In 
Fig.~\ref{fig:g28-g29-vs-b} we show, as an example, $G_{28}$ and $G_{29}$.
Note that the self-coupling is the dominant coupling.

\begin{figure}[t!]
\begin{center}
\epsfysize=63mm
\epsfxsize=70mm
\vspace{0.1cm}
\epsfbox{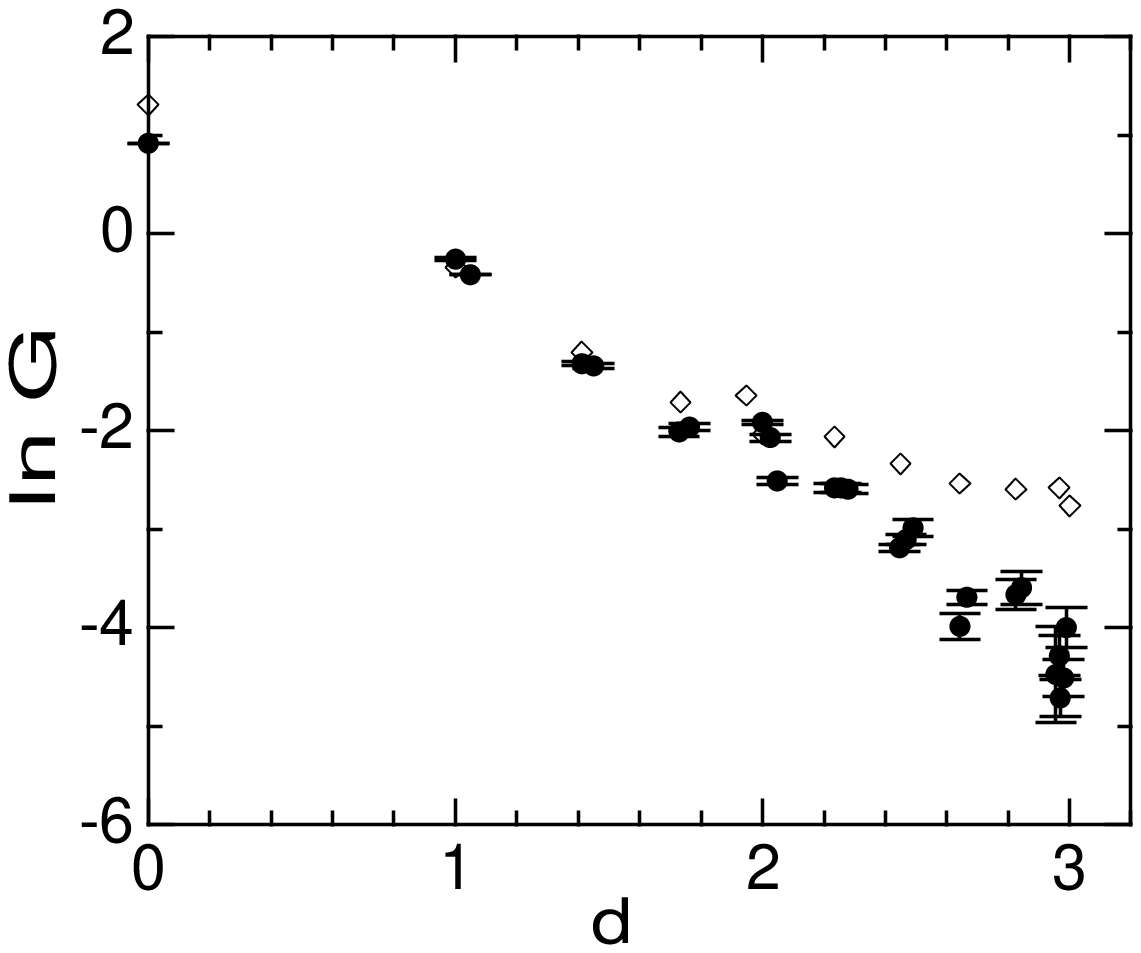}
\epsfysize=63mm
\epsfxsize=70mm
\epsfbox{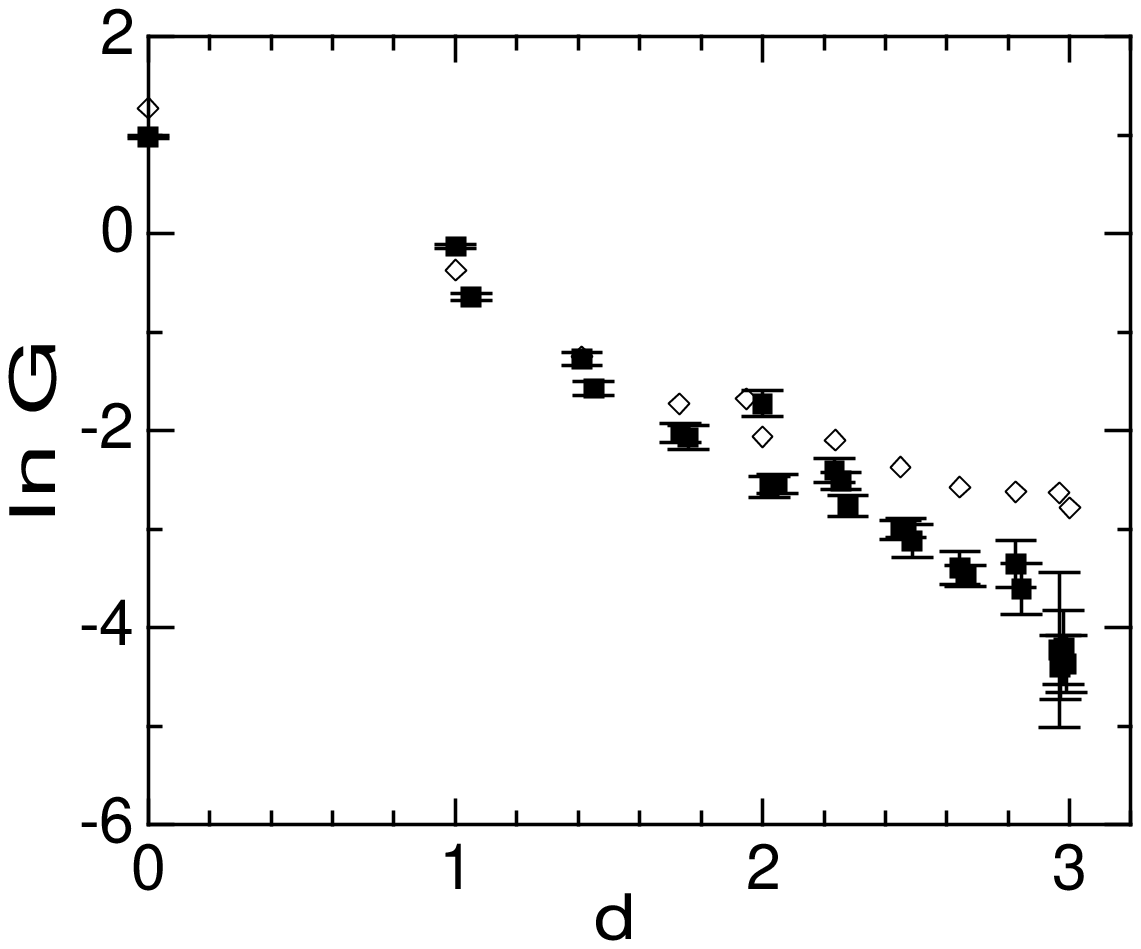}
\end{center} 
\vspace{-1.5cm}
\caption{\it The two-point coupling of monopole currents $k(s,\mu)$ and 
$k(s^\prime,\mu)$ as a function of 
distance $d=\sqrt{\sum_\mu(s_\mu-s^\prime_\mu)^2}$ in full and quenched QCD. The symbols are:
full QCD $(${\Large $\bullet$}$)$, quenched QCD $(\blacksquare)$ and Coulomb
propagator $(\lozenge)$.}
\label{fig:comp}
\vspace*{-0.7cm}
\end{figure}

To shed some more light on the dynamics of the monopoles we have looked at the
coupling $G$ of two ($n=1$) parallel monopole currents, 
$k(s,\mu)$ and $k(s^\prime,\mu)$,
as a function of their distance $d=\sqrt{\sum_\mu(s_\mu-s^\prime_\mu)^2}$.
In Fig.~\ref{fig:comp} we show $G$ together with the Coulomb propagator. We
see that at distances $d \gtrapprox 2$ the interaction becomes weaker than
Coulomb in both full and quenched QCD. This is a result of the screening 
effect discussed earlier on. It comes as a surprise though that we see no 
difference between full QCD and the quenched theory. We would have expected 
the screening effect to be considerably stronger in full QCD.

Our task for the future is to check for scaling of the effective monopole 
action, as this has already successfully been done in the quenched theory.

Furthermore, it would be useful to cast the action into the form
\begin{equation}
S(k)= S_{\rm Coulomb}+S_{\rm self}+S_{\rm 4-point}+S_{\rm 6-point},
\nonumber
\end{equation}
to make contact with the dual Ginzburg-Landau theory, and the string model,
and determine their parameters. 
This would allow us to do truly quantitative, analytic 
calculations~\cite{Kato}.

\section{CONCLUSIONS}

We have had a first look at the dynamics of color magnetic monopoles
in full QCD. We found striking differences between full QCD and the quenched 
theory. 

How can this be understood? Monopoles are at least partly induced by 
instantons~\cite{hart,bs}. The fermion determinant causes instantons
and anti-instantons to attract each other, with a force that increases with
decreasing quark mass. The effect is that isolated instantons are suppressed,
and instantons and anti-instantons are forced to form (overlapping) pairs.
This will increase the density of (anti-)instantons, and consequently the
density of monopoles.

\section{ACKNOWLEDGMENTS}
This work is partially supported by INTAS grant 00-00111.
H.I. is supported by a postdoctoral fellowship of the Japan Society
for the Promotion of Science JSPS. She acknowledges the hospitality of the
Humboldt-Universit\"at zu Berlin. T.S. acknowledges financial support 
from a JSPS Grant-in-aid for Scientific Research (B)  (No. 11695029).

\end{document}